# MarsCast: Transfer Learning of AI Weather Foundation Models to Planetary Atmospheres

M.L. Carroll[1α], J. Li[1,2], S.D. Guzewich[1], G.Villanueva[1], J.A. Caraballo-Vega[1], M.J. Frost[1,2]

1) NASA Goddard Space Flight Center
2) ADNet Systems, Inc.
α Corresponding author Mark.Carroll@nasa.gov

## Abstract

We investigate the transferability of Earth weather foundation models to planetary atmospheres by adapting the GraphCast graph neural weather forecasting model to Mars. While GraphCast achieves state-of-the-art performance for terrestrial forecasting, its applicability to non-Earth environments remains unexplored. Using the Mars Climate Database (MCD), which provides global atmospheric fields across vertical altitude levels (similar to Earth pressure levels), we evaluate zero-shot and fine-tuned GraphCast predictions of Martian temperature and wind fields. Zero-shot forecasts produce a surprisingly accurate depiction of current conditions but fail to reproduce diurnal variability and rapidly decay toward climatological mean states. To address this limitation, we fine-tune GraphCast using MCD variables and top-of-atmosphere solar radiation forcing while holding humidity constant. Fine-tuning enables rapid learning of Martian thermal variability. Within as few as 10 training epochs, the model begins to capture the diurnal cycle and forecasts up to 10 days reproduce seasonal and vertical temperature structure. Prediction quality improves with training sample size and exhibits sensitivity to seasonal initialization. These results demonstrate that Earth-trained AI weather models can be adapted to simulate Martian atmospheric dynamics, providing a pathway toward rapid planetary weather prediction to support mission operations, dust storm risk mitigation, and future human exploration.

## 1. Introduction

Recent advances in machine learning–based weather prediction have demonstrated that foundation models trained on terrestrial global atmospheric reanalysis datasets can rival traditional numerical weather prediction (NWP) systems while offering orders-of-magnitude reductions in computational cost (Bi et al., 2023; Bodnar et al., 2025; Lam et al., 2023). GraphCast, a graph neural network (GNN) weather forecasting model, represents atmospheric states on a geodesic grid and learns spatiotemporal evolution through message passing across the spherical domain (Lam et al., 2023). By learning atmospheric dynamics directly from data, such models offer a fundamentally different paradigm for weather prediction.

The idea behind foundation models in general is that they form a base upon which other models can be built through fine tuning. There are several recent efforts to generate AI based weather models for Mars largely from scratch (Roy et al., 2026; Sagar Uprety et al., 2025; Singh et al., 2026) none of these work with existing models for fine tuning. Fine-tuning AI weather foundation models beyond Earth provides an opportunity to evaluate their ability to generalize across fundamentally different atmospheric regimes and offers a potential pathway toward rapid atmospheric prediction for planetary science and exploration. In addition, the computational cost for fine-tuning is substantially lower than that for training a model from scratch.

Mars presents a compelling and scientifically rich testbed for this investigation. The Martian atmosphere is thin and $CO_2$-dominated, with surface pressure roughly two orders of magnitude lower than Earth's. Atmospheric dynamics are governed by strong diurnal temperature variability, seasonal mass exchange between polar caps and the atmosphere, pronounced topographic forcing from features such as Tharsis and Hellas, and radiative heating driven by suspended dust aerosols. These processes produce thermal tides, baroclinic waves, regional dust storms, and occasionally planet-encircling dust events. Accurate prediction of Martian atmospheric conditions is critical for mission planning, entry–descent–landing (EDL) safety, surface operations, and future human exploration.

Numerical simulation of the Martian atmosphere is traditionally performed using Mars Global Climate Models (MGCMs), which solve the governing equations of atmospheric motion, thermodynamics, and radiative transfer (Forget et al., 1999; Haberle et al., 1993; Richardson et al., 2007). Although physically comprehensive, MGCM simulations are computationally intensive and not well suited for rapid forecasting or ensemble-based uncertainty quantification. AI-based atmospheric prediction offers a complementary approach capable of producing rapid forecasts once trained.

A key open question is "Can weather foundation models trained on Earth atmospheric reanalysis data be transferred to planetary environments with distinctly different atmospheric composition and physics?". Terrestrial models implicitly encode assumptions about atmospheric composition, radiative balance, circulation regimes, and boundary forcing that differ substantially from Martian conditions. Evaluating transferability provides insight into the degree to which atmospheric dynamics are universal versus planet-specific.

In this study, we investigate the transferability of the GraphCast weather foundation model to Mars using atmospheric fields from the Mars Climate Database (MCD). We first evaluate zero-shot autoregressive forecasts to assess whether the Earth-trained model can reproduce Martian atmospheric variability. We then develop a targeted fine-tuning strategy focused on temperature and wind variability and solar forcing to enable prediction of Martian thermal dynamics.

This work addresses the following science and modeling questions:

1. **Transferability of atmospheric foundation models** — Can an Earth-trained graph neural net weather model reproduce Martian atmospheric variability without retraining?
2. **Failure modes in zero-shot planetary prediction** — What physical processes are not captured when terrestrial models are applied directly to Mars?
3. **Learning planetary thermal dynamics** — Can targeted fine-tuning enable the model to learn Martian diurnal and seasonal temperature variability?
4. **Vertical and seasonal structure** — Can the adapted model reproduce temperature evolution across vertical pressure levels and seasonal cycles?

5. **Data efficiency and learning behavior** — How sensitive is model performance to training duration and sample size?

By addressing these questions, this study evaluates the feasibility of adapting Earth weather foundation models for planetary atmospheric prediction and explores their potential role in supporting Mars science and exploration.

## 2. Data and Computational Resources

### 2.1 ERA5 Context and Pretraining Domain

GraphCast was originally trained on ERA5 reanalysis data, which provides globally consistent atmospheric fields derived from data assimilation of observations and numerical weather prediction. ERA5 spans multiple decades with hourly resolution and includes three-dimensional atmospheric state variables. The model therefore encodes atmospheric dynamics representative of Earth's radiative balance, thermodynamics, and circulation regimes. Because GraphCast was pretrained on ERA5 temperature distributions, Mars temperature fields were rescaled to the ERA5 dynamic range to maintain numerical consistency and avoid activation saturation within the pretrained network. The transformation is

$$T_{scaled} = \frac{T_{Mars} - T_{Mars}^{min}}{T_{Mars}^{max} - T_{Mars}^{min}}\left(T_{ERA5}^{max} - T_{ERA5}^{min}\right) + T_{ERA5}^{min}$$

Where

$T_{Mars}$ is the original Mars variable

$T_{Mars}^{min}$ and $T_{Mars}^{max}$ are the minimum and maximum Mars variable

$T_{ERA5}^{min}$ and $T_{ERA5}^{max}$ are the corresponding minimum and maximum variable from the ERA5 dataset

$T_{scaled}$ Is the rescaled Mars variable used as model input.

### 2.2 Mars Climate Database (MCD)

The Mars Climate Database provides global atmospheric fields derived from Mars Global Climate Model simulations and data assimilation products (Millour et al., 2024). The dataset offers global coverage, long temporal records, and vertically resolved atmospheric structure with variables that are analogous to the variables in ERA5 suitable for machine learning applications.

Variables used in this study include:

- 2-meter air temperature
- Atmospheric temperature at 13 pressure levels (50–1000 Pa)
- Surface temperature
- Top-of-atmosphere (TOA) solar radiation forcing

Temperature and wind fields were min–max scaled to match the ERA5 temperature range used during GraphCast pretraining. Humidity variable was held constant at the global mean value following z-score normalization in order to isolate temperature variability and reduce dimensional complexity during initial adaptation. Mars was treated as entirely land by setting the land–sea mask to unity. The seasonal cycle was represented using solar longitude (Ls), discretized into 360 values representing one Martian year. Local time sampling was performed at 00, 06, 12, and 18 hours to capture diurnal variability.

Table 1. Shows the inputs for GraphCast and the corresponding inputs used from the MCD to fine-tune GraphCast for Mars. Additionally, the 13 vertical altitude levels that were used to simulate the 13 vertical levels from GraphCast are described.

| **GraphCast Inputs** | **MCD Dataset** | |
|---|---|---|
| 1° x 1° latitude and longitude grid | 5.625° x 3.75° bilinear interpolation to 1° x 1° | |
| 13 vertical pressure levels | 13 vertical altitude levels corresponding to 50, 100, 200, 250, 300, 400, 500, 600, 700, 850, 925, and 1000 Pa | |
| ***Variables*** | | |
| 2-m temperature | 2-m temperature | 2D Variables |
| Mean sea level pressure | Surface pressure | |
| 10-m u component of wind | 2-m West-to-East wind component | |
| 10-m v component of wind | 2-m South-to North wind component | |
| Sea surface temperature | Computed from temperature and pressure | |
| Temperature | Temperature | 3D Variables |
| U component of wind | West-to-East wind component | |
| V component of wind | South-to North wind component | |
| Vertical velocity | Downward vertical wind component | |
| Geopotential | Computed from temperature and pressure | |
| Specific humidity | Dust mass mixing ratio | |
| Total precipitation | Total precipitation set to 0 | Static and Forcing |
| Land-Sea mask | Land-Sea mask set to 1 (no ocean) | |
| Geopotential at surface | Computed from temperature and pressure at surface | |

## 2.3 Computational Resources

All experiments were conducted using GPU-accelerated high-performance computing resources at NASA Goddard Computational and Information Science and Technology Office. Training and inference were performed using PyTorch with mixed-precision acceleration. Fine-tuning leveraged the pretrained GraphCast weights and required substantially reduced computational cost compared to training from scratch. Typical fine-tuning experiments completed within hours on a single modern GPU, while inference produced global forecasts in seconds, highlighting the potential for rapid operational deployment.

## 3. Methods

### 3.1 GraphCast Architecture

GraphCast is a graph neural network–based weather forecasting system that represents the atmosphere on a geodesic grid to preserve spherical geometry (Lam et al., 2023). Each grid cell is represented as a node containing atmospheric state variables, while edges encode spatial relationships between neighboring cells. This graph representation enables physically consistent propagation of information across the globe.

The model employs an encoder–processor–decoder architecture. The encoder transforms atmospheric state variables into latent embeddings. The processor applies iterative message passing across the graph, enabling spatial information exchange and representation of large-scale atmospheric dynamics. The decoder projects latent states back into physical variables to produce forecasts. This architecture allows efficient modeling of global atmospheric evolution while preserving spatial structure and scale interactions.

### 3.2 Zero-Shot Forecast Experiment

To evaluate transferability, GraphCast was applied autoregressively to Martian atmospheric states without Mars-specific training. Forecast skill was evaluated by examining temporal stability, spatial gradients, and diurnal variability.

Zero-shot predictions rapidly decayed toward climatological mean conditions and exhibited a complete absence of diurnal variability. Thermal gradients weakened with forecast lead time, indicating that Earth-trained dynamics do not generalize to Martian radiative forcing and atmospheric structure ( Figure 1).

### 3.3 Fine-Tuning Strategy

To enable prediction of Martian atmospheric variability, the adaptation problem was reduced to temperature and wind dynamics.

Model inputs included:

- 2-meter temperature
- 10-meter wind
- Temperature, W-E wind and S-N wind at 13 pressure levels
- TOA solar radiation forcing
- Static surface and topographic influences

Training was performed using the AdamW optimizer with a learning rate of $1 \times 10^{-6}$. Mean squared error (MSE) loss was used to optimize temperature predictions. All network components, including encoder, processor, and decoder layers, were unfrozen to allow adaptation of learned representations. A small learning rate was chosen to preserve pretrained spatial representations while enabling Mars-specific thermal dynamics to be learned.

### 3.4 Training Sample Construction

Training samples were constructed to capture diurnal evolution. For each seasonal date, sequential triplets were generated:

- (T00, T06 → T12)
- (T06, T12 → T18)
- (T12, T18 → T00)
- (T18, T00 → T06)

This formulation enables the model to learn diurnal heating and cooling driven by solar forcing.

### 3.5 Training Procedure and Sensitivity Experiments

Fine-tuning experiments evaluated sensitivity to training duration and sample size. Training was allowed to continue to ~1000 epochs with checkpoints saved every two epochs. Loss decreased rapidly during early epochs, with the model beginning to capture diurnal variability after approximately 10 epochs (Figure 2). Autoregressive rollouts were performed using the model checkpoints with those results compared to dates that were held out of the training.

Sample size experiments were conducted using subsets of 30, 100, and 360 days to assess data efficiency. These experiments quantify how quickly planetary thermal dynamics can be learned.

### 3.6 Forecast Generation and Evaluation

Forecasts were generated autoregressively for lead times up to 10 days. Model performance was evaluated for:

- 2-meter temperature
- vertical temperature profiles
- seasonal initialization states
- vertical and horizontal winds

Evaluation focused on the model’s ability to reproduce diurnal cycles, vertical structure, and seasonal variability.

## 4. Results

### 4.1 Zero-Shot Performance

Zero-shot forecasts produce discernible patterns for both temperature and winds, however they fail to reproduce diurnal variability and collapse toward climatological equilibrium within a few time steps (Figure 1) demonstrating that terrestrial atmospheric dynamics do not directly generalize to Mars.

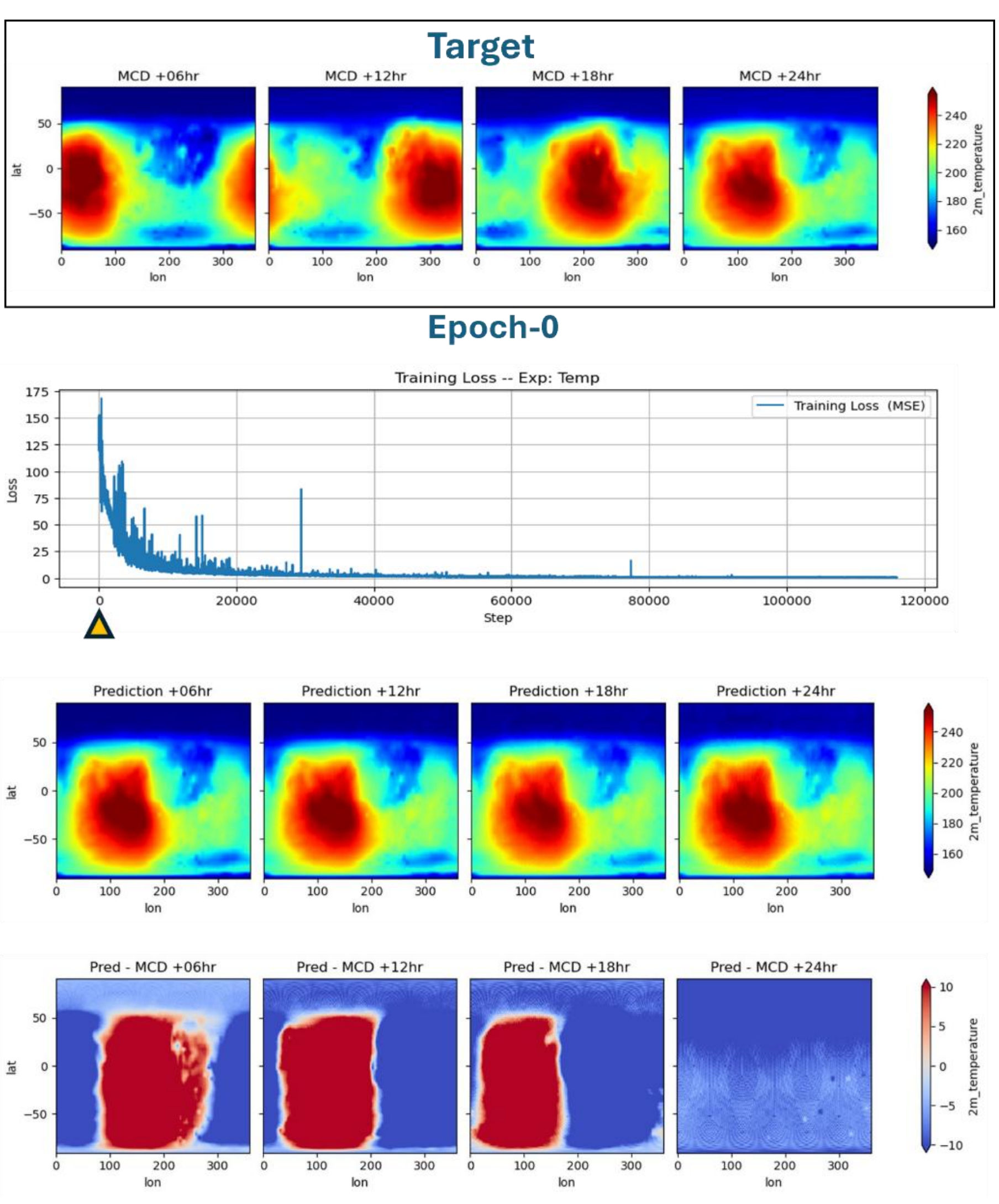

Figure 1. 24-hour forecast (6 hour time steps) generated from GraphCast with no fine tuning (i.e. zero-shot). The target (Mars Climate Database) is shown at the top, training loss shown in the middle, prediction in the third row and the difference between the target and the prediction in the bottom row.

### 4.2 Rapid Learning During Fine-Tuning

Fine-tuning GraphCast using MCD data (henceforth called MarsCast) enables rapid learning of Martian atmospheric dynamics. Within approximately 10 epochs, the model begins to capture the diurnal heating and cooling cycle (Figure 2). This demonstrates that the atmospheric dynamics learned by the GraphCast model are transferrable to the Mars atmosphere, the fine tuning facilitates refining of the magnitudes of the variables and learning the nuances of the Martian atmosphere. Continued training improves both amplitude accuracy and spatial coherence (Figure 2). The model was allowed to train for ~1000 epochs with early stopping turned off. The fine-tuned model reproduces diurnal temperature variability consistent with solar forcing and maintains realistic spatial gradients.

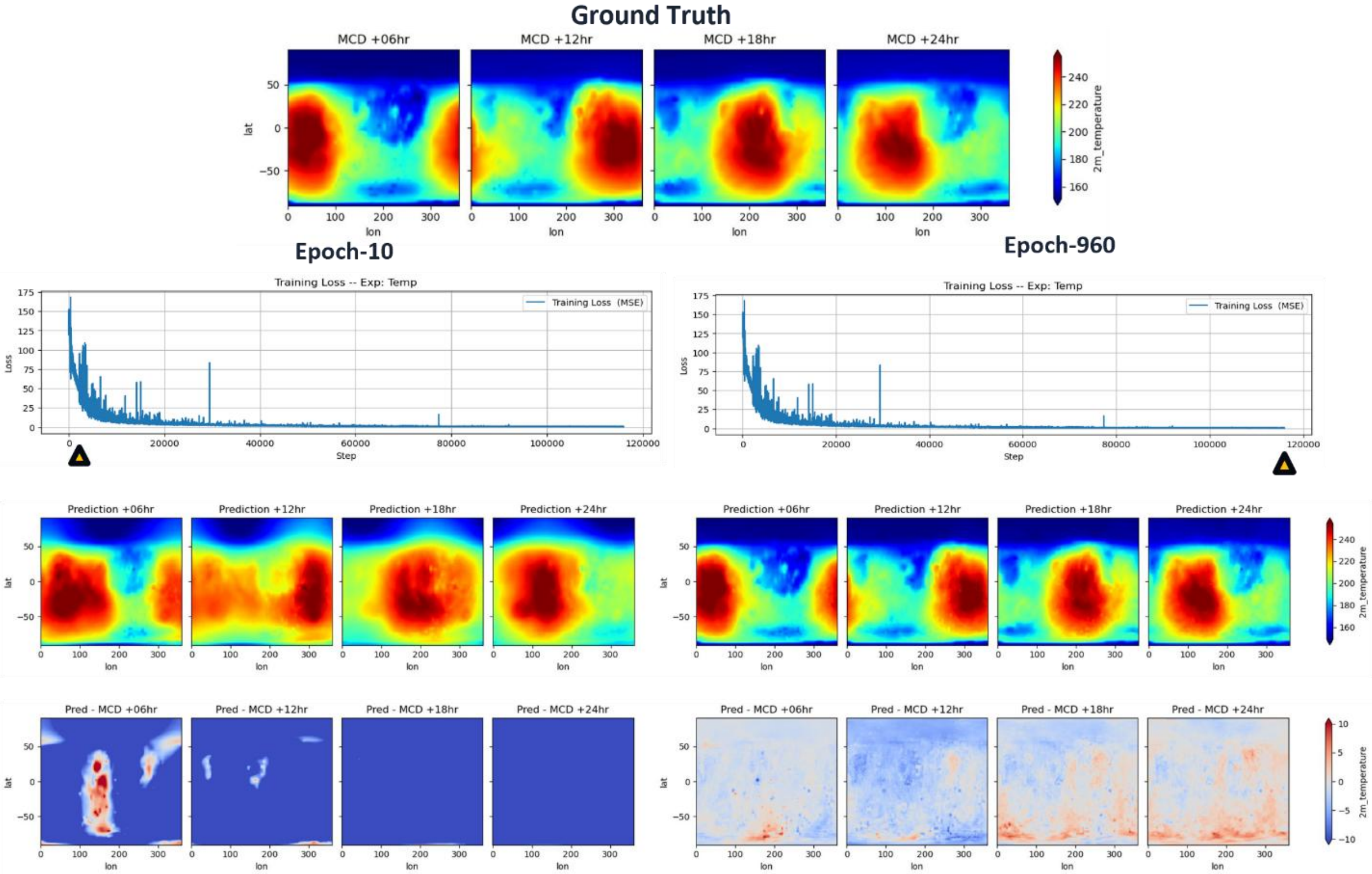

Figure 2. A 24-hour forecast for temperature at 2 m after fine-tuning GraphCast for only 10 epochs (left side) with a small amount (30 days randomly selected) of training. The diurnal cycle is already starting to appear. After training for ~1000 epochs (right side) the diurnal cycle is fully developed, and the errors have been reduced to less than +/- 2 degrees in most locations.

This fine-tuned model was used in all further experiments, model architecture and checkpoints can be found in GitHub https://github.com/nasa-nccs-hpda/GraphCastMars and Hugging Face nasa-cisto-data-science-group/graphcast-mcd-wind-temperature. We used a checkpoint with the lowest validation RMSE in the remaining experiments because we found this to produce the best results.

### 4.3 Vertical Temperature Structure

The MarsCast model produces outputs at 13 vertical levels, as it is built on the GraphCast model which also has 13 levels (see table 1 for more details on pressure levels included). Figure 3 compares predicted vertical temperature profiles with the MCD at the location near NASA's Phoenix Mars Lander (68.22ºN, 234.25ºE) for leading time ranging from 6 to 78 hours. MarsCast successfully reproduces the overall vertical thermal structure throughout the forecast period, with particularly good agreement in the middle atmosphere (approximately 250-800 Pa). The largest deviations occur in the lower and upper atmosphere (pressure near ~850 Pa, and below ~200 Pa), and these discrepancies become more pronounced with increasing lead time. Nevertheless, the predicted profiles preserve the major vertical temperature gradients and near surface thermal structure, indicating that MarsCast maintains skill in representing the vertical temperature distribution over multi-day forecasts. This performance is consistent with the performance seen in the evaluation of the native GraphCast model (Lam et al., 2023).

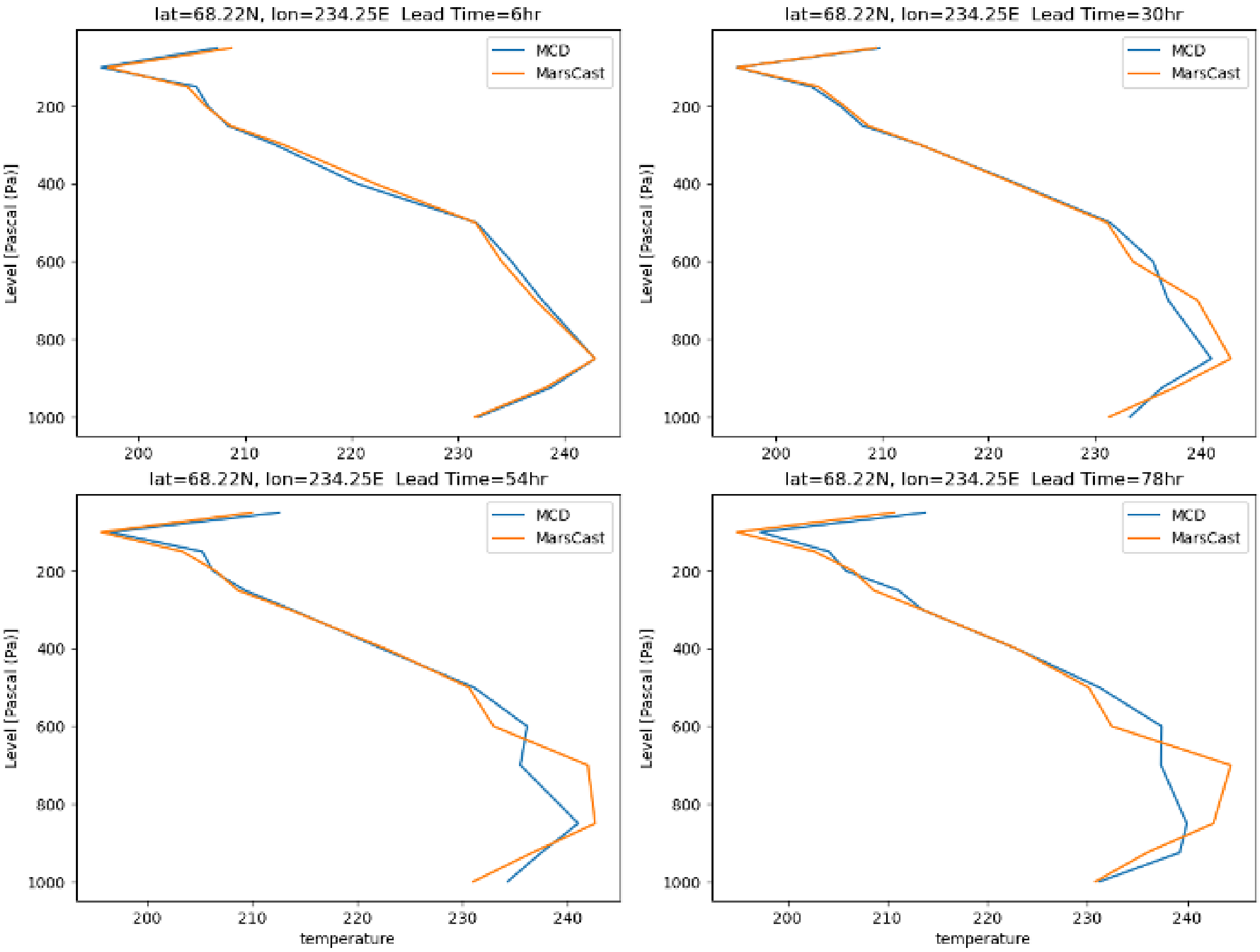

Figure 3. Vertical temperature profiles predicted by MarsCast (orange) and the Mars Climate Database (MCD) (blue) at the Phoenix Mars Lander site (68.22°N, 234.25°E) for forecast lead times of 6, 30, 54, and 78 hours.

### 4.4 Seasonal Performance

The fine-tuned MarsCast model was evaluated using 96-hour forecasts initialized at four representative seasons (Ls = 60°, 150°, 240°, and 330°). Figure 4 demonstrates that forecast skill is maintained across all seasonal initializations, indicating that the model successfully captures the seasonal dependence of the Martian atmosphere. Surface temperature and lower-atmosphere temperatures (700 Pa) are reproduced with high fidelity throughout the forecast period, while forecast errors increase in the upper atmosphere (50 Pa), consistent with the greater dynamical variability at higher altitudes (Figure 5). Overall, these results suggest that the fine-tuned model effectively represents both the seasonal forcing and the three-dimensional thermal dynamic structure of the Martian atmosphere, with the strongest performance in the lower atmosphere.

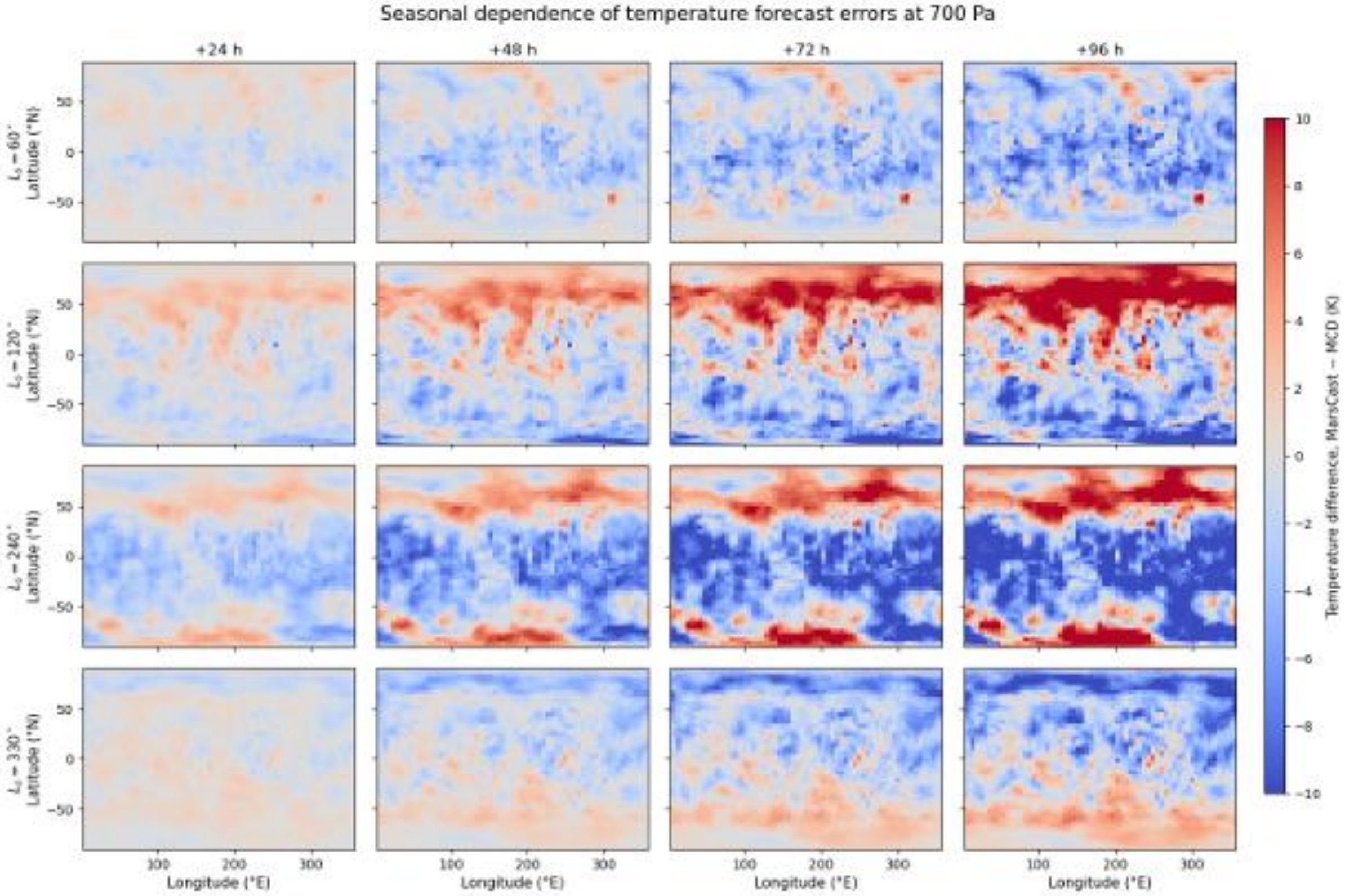

Figure 4. Seasonal dependence of MarsCast 700-Pa temperature forecast errors over a 96-hour forecast period. Rows correspond to forecasts initialized at four representative solar longitudes Ls=60o, 120o, 240o, and 330o, while columns show forecast lead times of 24, 48, 72, and 96 hours. Colors indicate the temperature difference between MarsCast and the MCD reference field (MarsCast − MCD), with negative values representing a cold bias and positive values a warm bias.

Additional figures showing seasonal and vertical performance can be seen in the supplemental figures section.

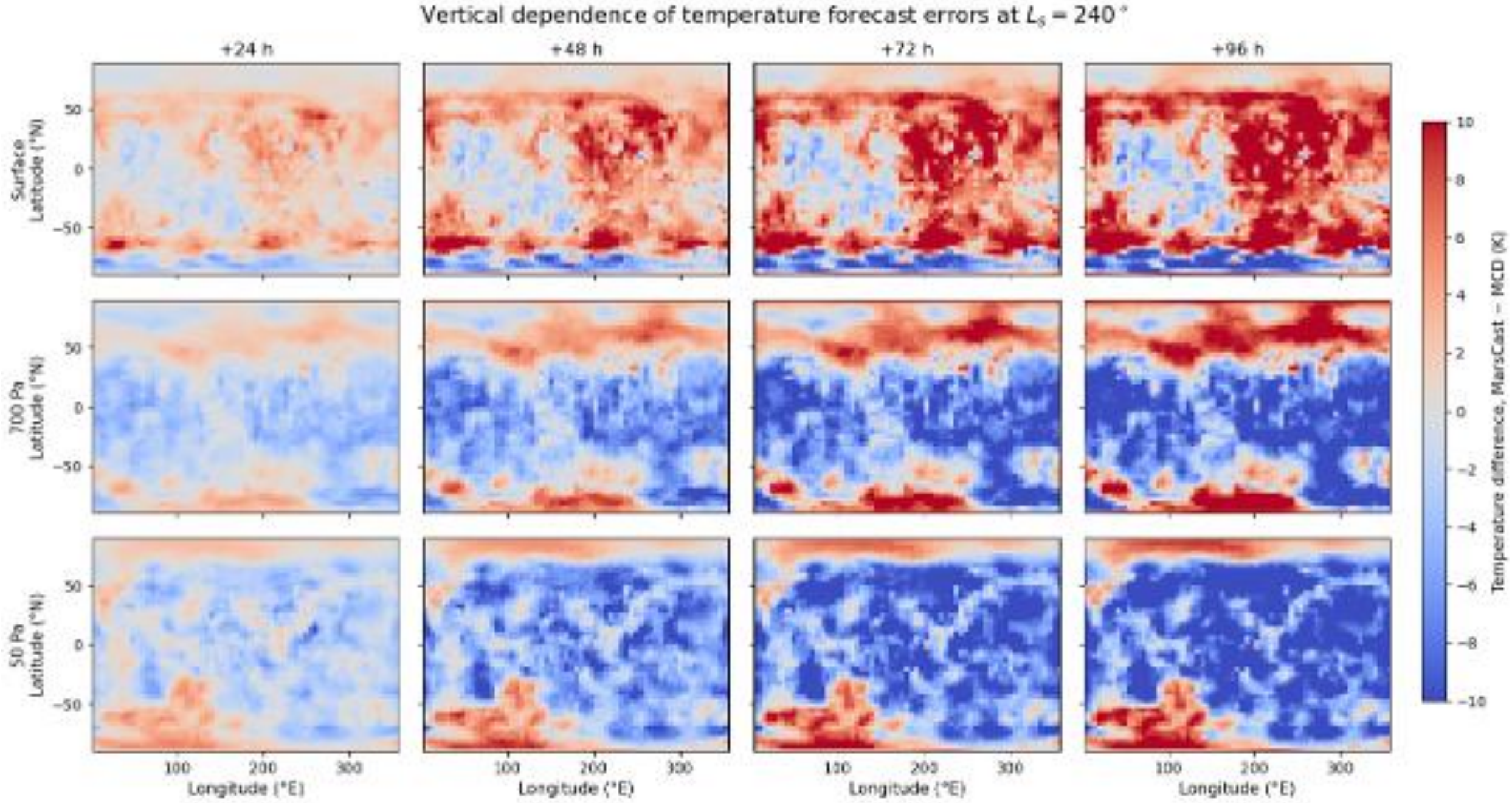

Figure 5. Vertical dependence of MarsCast temperature forecast errors for a 96-hour forecast initialized at Ls=240o. Rows show the surface (2-m temperature), 700 Pa, and 50 Pa temperature fields, while columns correspond to forecast lead times of 24, 48, 72, and 96 hours. Colors indicate the temperature difference between MarsCast and the MCD reference field (MarsCast − MCD), with positive values indicating a warm bias and negative values indicating a cold bias.

### 4.5 Wind performance

In the initial experiments only temperature was predicted with all other variables including winds held to the global mean. For the final experiment both meridional and zonal winds were included during fine-tuning, enabling MarsCast model to jointly predict meridional and zonal winds in addition to temperature (Figure 5). Figure 6 demonstrates that the fine-tuned model successfully reproduces the large-scale temperature distribution and near-surface wind circulation over a complete Martian diurnal cycle. The predicted winds closely follow those of the MCD, with wind speed differences generally remaining within +/- 4 m/s for at least the first 48 hours of the prediction. Temperature errors are small over most of the domain, typically within +/- 2 K, and are approximately centered around zero, indicating little systematic warm or cold bias.

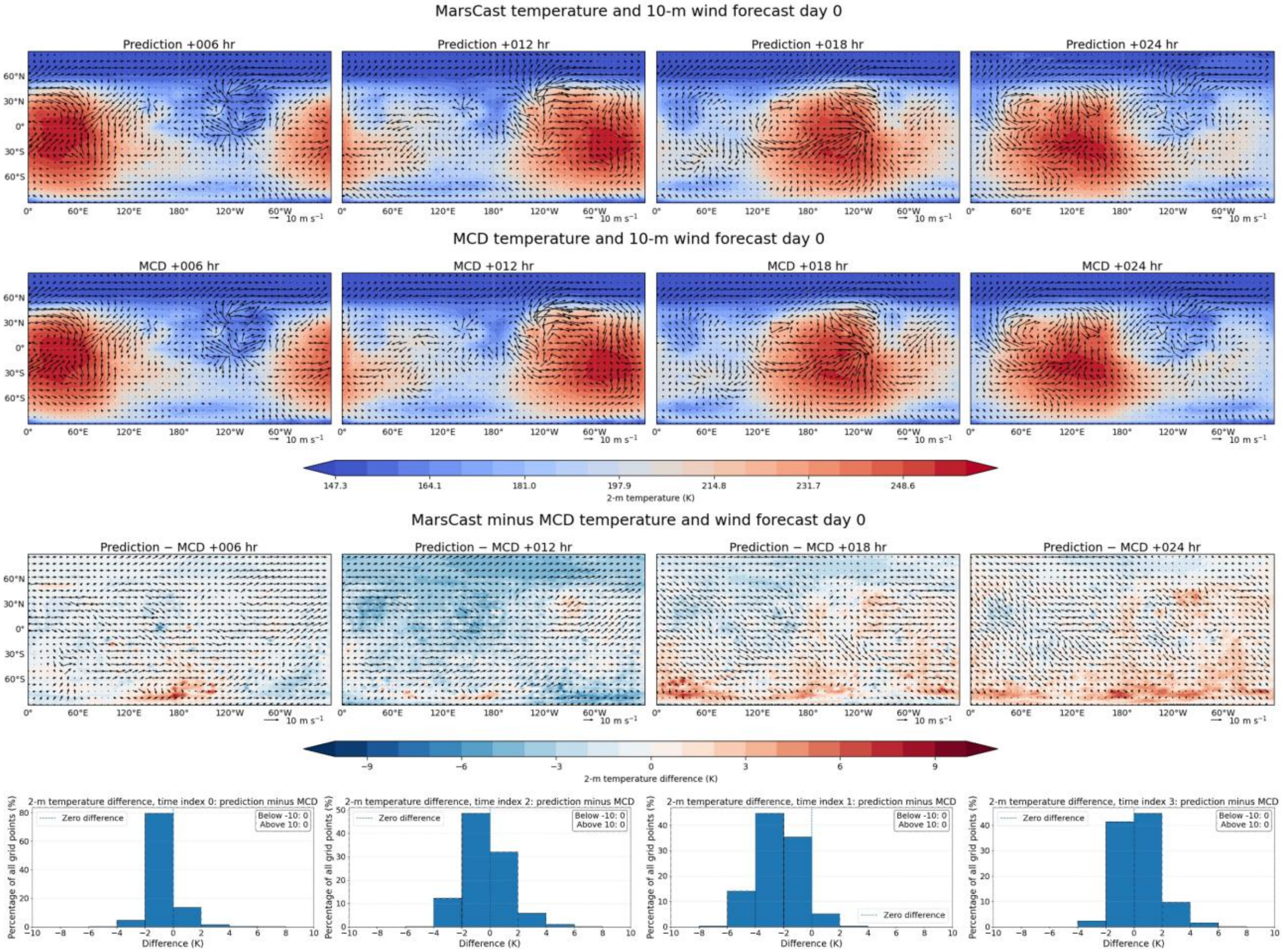

Figure 6 Comparison of MarsCast predictions and MCD reference fields over one Martian diurnal cycle. The top row shows predicted near-surface (10 m) temperature (color shading) and horizontal wind vectors, while the middle row shows the corresponding MCD fields. The bottom row presents the temperature difference (MarsCast − MCD) with the difference wind vectors overlaid. The accompanying histograms summarize the distribution of temperature errors for each forecast time. MarsCast accurately reproduces the large-scale temperature patterns and near-surface wind circulation, with temperature differences generally within ±2 K and wind speed differences within approximately ±4 m $s^{-1}$ during the early forecast period.

## 5. Discussion

The performance of zero-shot run of GraphCast initialized with MCD highlights the value of Foundation Models as a "base" upon which other models can be built. The model, trained only on Earth atmosphere data, captured the patterns from the initialization but was unable to produce the diurnal cycle. This highlights the importance of planetary-specific radiative forcing, atmospheric composition, and thermal inertia in shaping atmospheric dynamics. Earth weather models encode dynamics appropriate for a dense, water-vapor-rich atmosphere and therefore fail to reproduce Mars' radiative balance and diurnal forcing.

Fine-tuning with a small amount of data, samples from just 30 days throughout the Martian year, and training for only 10 epochs, the GraphCast model begins to learn the patterns and magnitudes of the variables. After ~300 epochs the model has stabilized somewhat but continues to improve all the way out to ~1000 epochs. The success of fine-tuning is largely because the model retains general spatiotemporal learning capacity while adapting to Mars-specific forcing conditions. This suggests that atmospheric foundation models learn transferable dynamical structures that can be adapted with modest retraining.

The results were successful not only for temperature but also for vertical and horizontal winds which demonstrates the feasibility of adapting Earth AI weather foundation models to planetary atmospheres and provide insight into the universality of learned atmospheric dynamics. A fundamental difference between Earth and Mars atmospheres lies in the amount of water both on the surface and in the atmosphere. For purposes of this example the surface was considered all land and the atmospheric humidity was held constant at the global mean, which is unrealistic. *In situ* observations indicate relative humidity is <10% for most of the sol, with brief excursions to higher values (occasionally even reaching saturation) just before dawn when air temperatures are lowest (Martínez et al., 2017; Polkko et al., 2023). Relative humidity at the surface appears highly dependent on surface properties, while being modulated by the global water cycle (e.g., Pál et al., 2019).

The intrinsic value of using an AI based weather foundation model on Mars is three-fold. 1) the time to forecast is orders of magnitude lower than running the full physics model. This could come into play in future operations on Mars (both human and robotic). 2) computational cost for the 10 day forecast from the foundation model is substantially lower: 1 GPU for ~2 minutes, to get a 1° resolution forecast compared to 1 CPU for 30 minutes to get a 5° resolution forecast from the physics model (i.e. Foundation Model is processing 25x more data in 2 minutes than the physics model is in 30 minutes). Reduced computation cost increases the options for running ensembles which could also improve the forecast. 3) GraphCast was trained at 0.25° spatial resolution and in this application it was run at 1° spatial resolution which is substantially finer than the 3.75° x 5.625° native spatial resolution of the MCD. The MCD was downscaled to 1° using bilinear interpolation for initialization of the foundation model and the outputs visually appear to capture some of the finer resolution details from the original model. This implies that a basic downscaling of the coarser resolution inputs can result in a more detailed forecast output.

The MarsCast model was run autoregressively to get a 10-day forecast at 6 hour time steps. For this example we are running historical data so we can compare the forecast to the MCD data for the forecasted days. The comparison shows the forecast degrades over time relative to the MCD, which is consistent with any forecast. Additional work needs to be done to quantify, including comparison with observations to quantify the degradation over time, but anecdotally the forecast looks good (by standards of public-facing terrestrial weather forecast accuracy) for at least 96 hours.

To understand the robustness of the model we initialized forecasts from many different days in all four seasons and produced 10-day forecasts for each (figures in supplemental materials). The results showed that MarsCast was correctly representing patterns consistent with changing seasons on Mars and maintained a consistent performance in the often variable Ls = 240° season.

Future work will extend the model to incorporate dust transport, radiative effects, and to validate predictions using orbital and surface observations. Integration with Mars mission planning frameworks and engagement with the Mars Exploration Program will enable operational applications.

## 6. Conclusion

The idea of a foundation model is that it is a base upon which additional models can be built or extended to other areas. This study demonstrates that an Earth trained AI weather foundation model (GraphCast) can be adapted to forecast Martian weather through targeted fine-tuning. Even the "zero-shot" result show a pattern that is relevant to the conditions on Mars. Fine-tuning an existing foundation model is a cost-effective way to generate a new model for a different task, in this case Mars weather. The resulting fine-tuned model, we called MarsCast, captures diurnal and seasonal variability in both temperature and winds and enables rapid atmospheric prediction, providing a pathway toward AI-augmented planetary weather forecasting.

## Supplemental materials

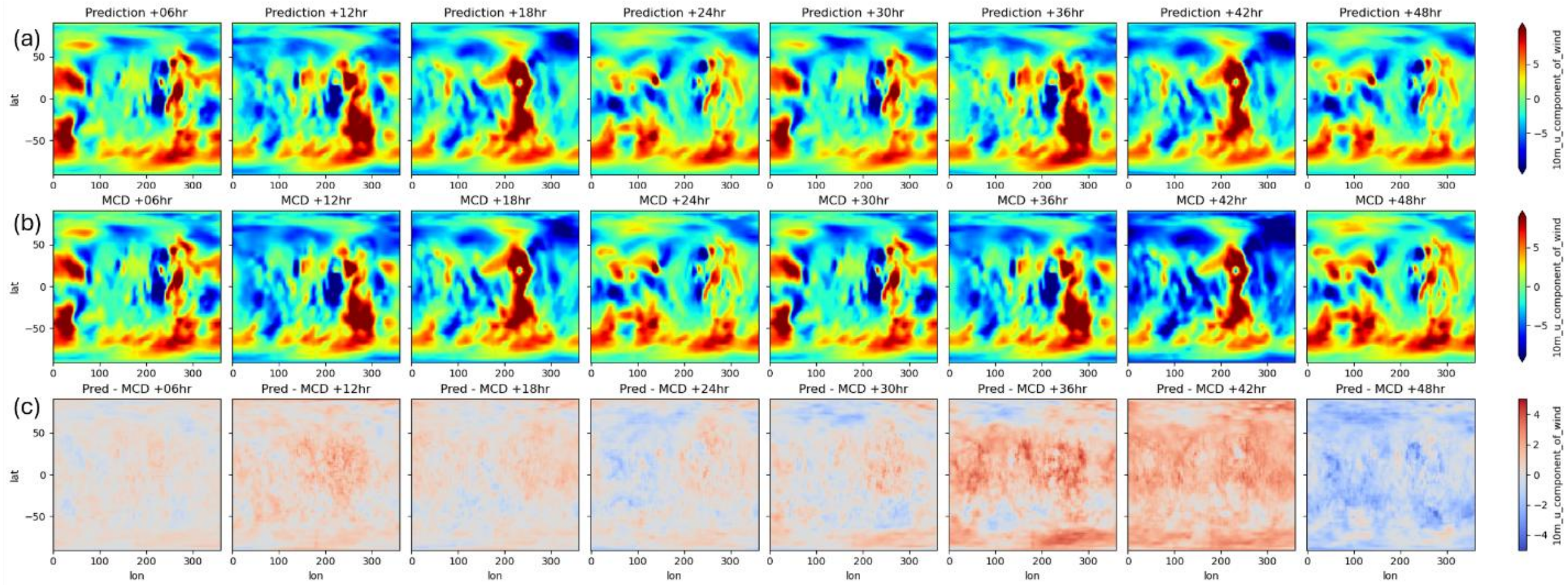

Figure S1: For 10-m zonal "u" wind component, comparison of (a) MarsCast predictions (b) MCD reference and (c) difference at for +06 hours to +48 hours, initialized for solar longitude (Ls = 60) in the third "spring" month.

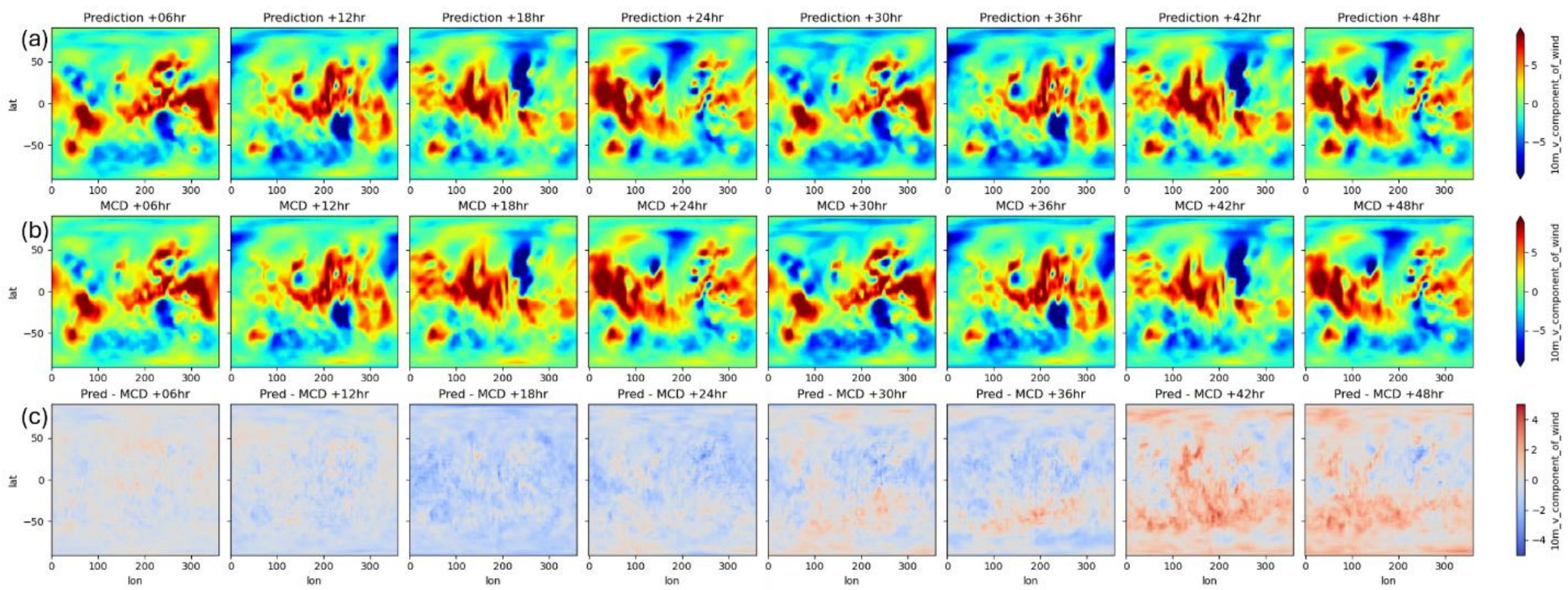

Figure S2: For 10-m meridional "v" wind component, comparison of (a) MarsCast predictions (b) MCD reference and (c) difference at for +06 hours to +48 hours, initialized for solar longitude (Ls) = 60 in the third "spring" month.

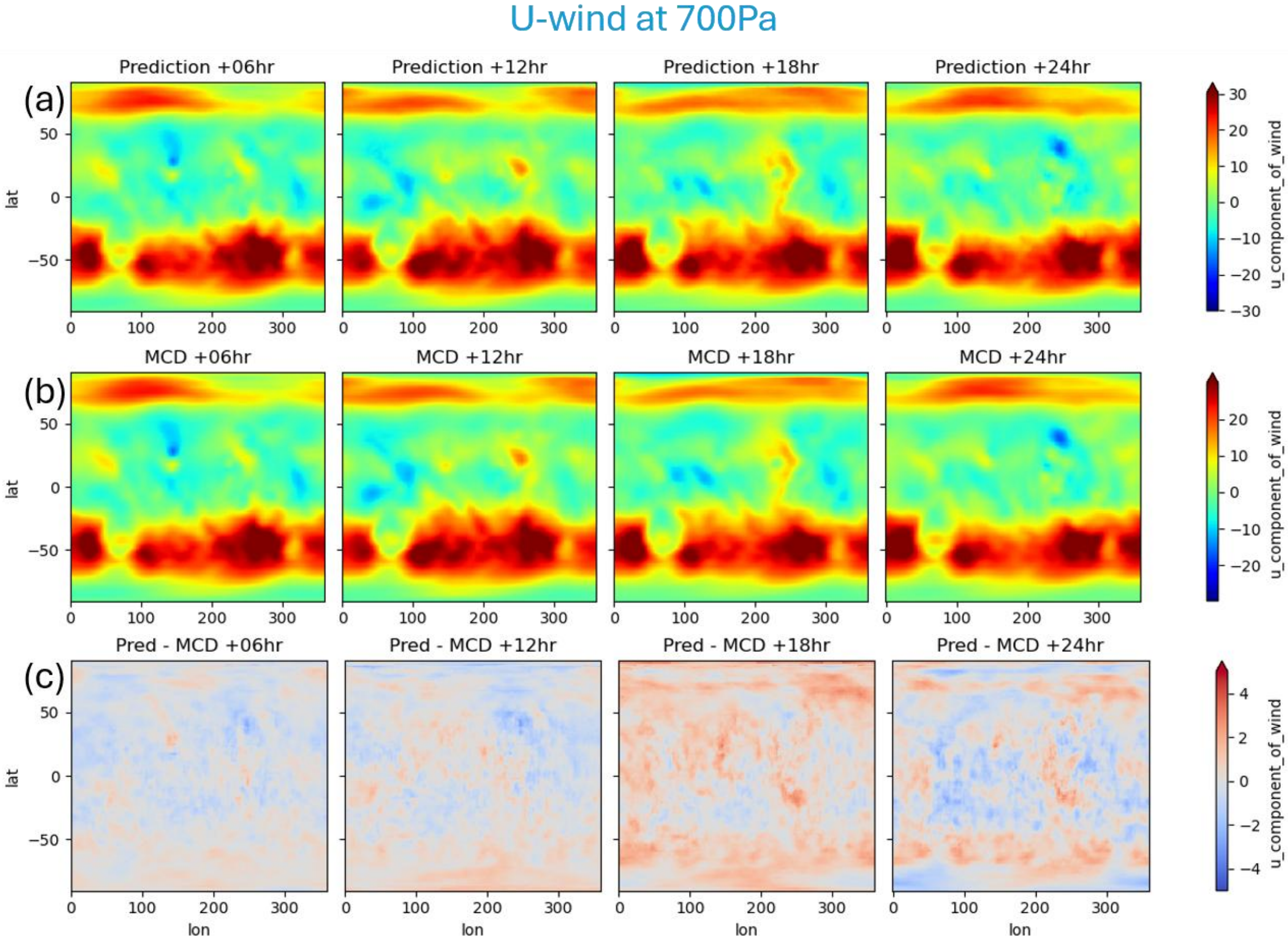

Figure S3: For 10-m zonal "u" wind component, comparison of (a) MarsCast predictions (b) MCD reference and (c) difference at for +06 hours to +48 hours, at 700Pa.

V-wind at 700Pa

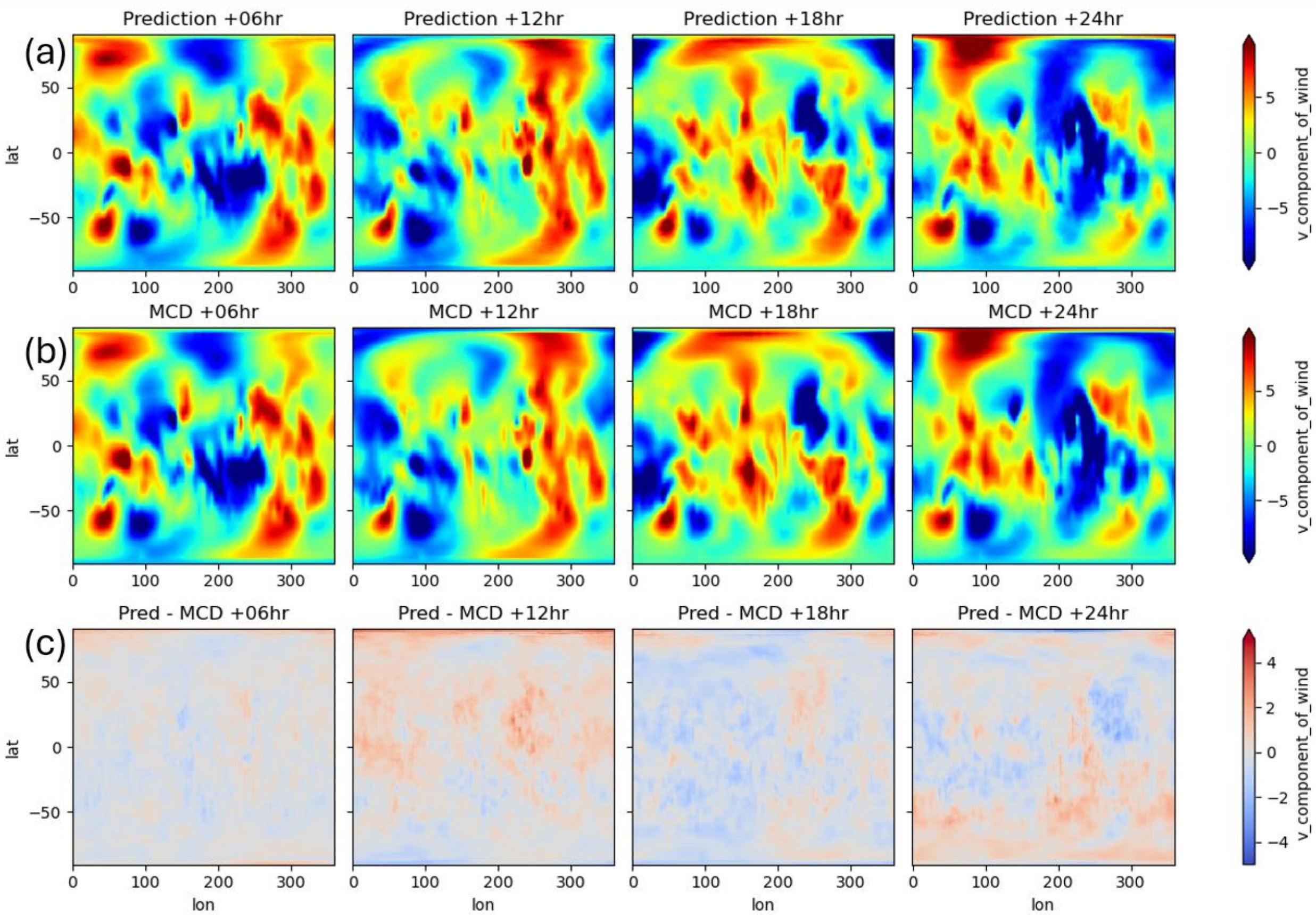

Figure S4: For 10-m meridional “v” wind component, comparison of (a) MarsCast predictions (b) MCD reference and (c) difference for +06 hours to +48 hours, at 700Pa

| ***Fine-Tune Configuration*** | |
| --- | --- |
| Input Data Standardization - MCD | 2m-temperature, temperature @ 13 z-levels (minmax scaled to ERA5 range) |
| Input Data Standardization - other | global mean (~0.0, after z-score normalization) |
| Forcing | Accumulated TOA solar radiation |
| Static | Land-Sea mask == 1 |
| Calendar Date | Choose 360 solar longitudes to represent seasonal cycle |
| Local Time | T00, T06, T12, T18 |
| Training Samples | For each date, four samples: (T00, T06, T12), (T06, T12, T18), (T12, T18, T00), & (T18, T00, T06) |
| Optimizer | AdamW |
| Learning Rate | 1e-6 |
| Loss Function | Mean Squared Error loss |
| Unfrozen components | Encoder, Processor, & Decoder |

Table S1: Configuration of the data standardization, adaptations to fit Mars to Earth measurements, and model parameters. The land-sea mask was set to 1 since Mars does not have water bodies, and the initial learning rate was small as the model is large.

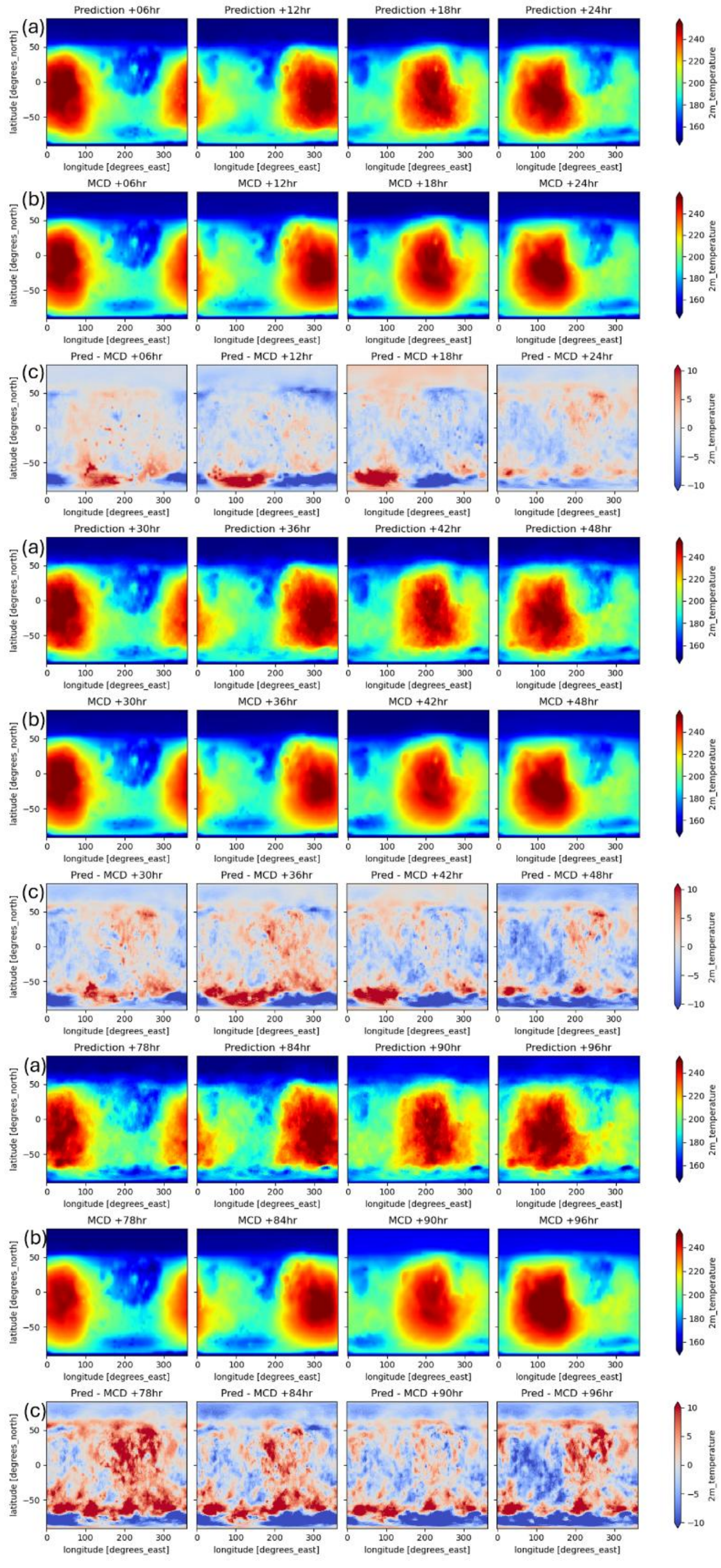

Figure S5: 2-m surface temperature, initialized at Ls = 330, T00 and T06 for (a) MarsCast Prediction, b) MCD reference and (c) difference for +06 to +24 hours (top), +30 to +48 hours (middle), and +54 to 72 hours (bottom)

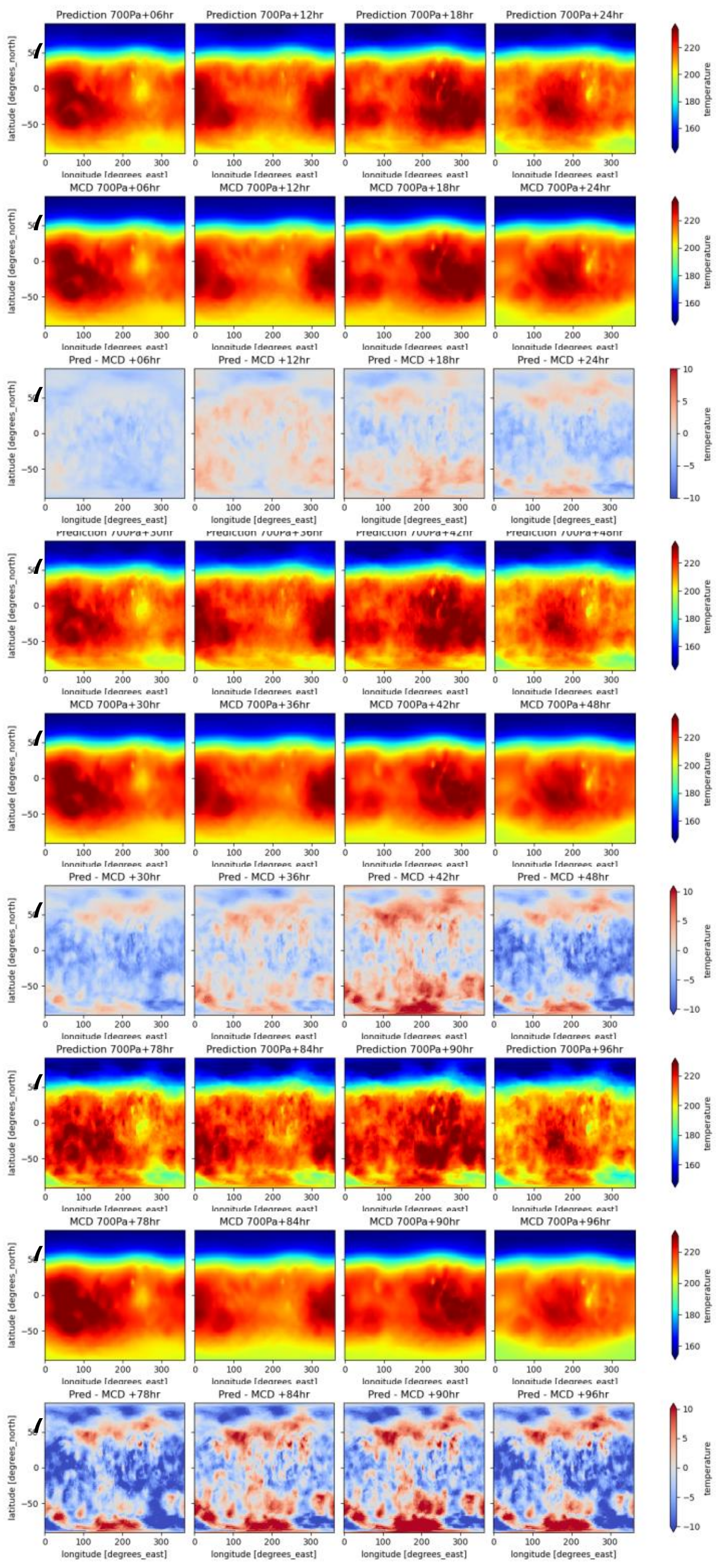

Figure S6: Temperature at 700Pa, initialized at Ls = 330, T00 and T06 for (a) MarsCast Prediction, b) MCD reference and (c) difference for +06 to +24 hours (top), +30 to +48 hours (middle), and +54 to 72 hours (bottom)

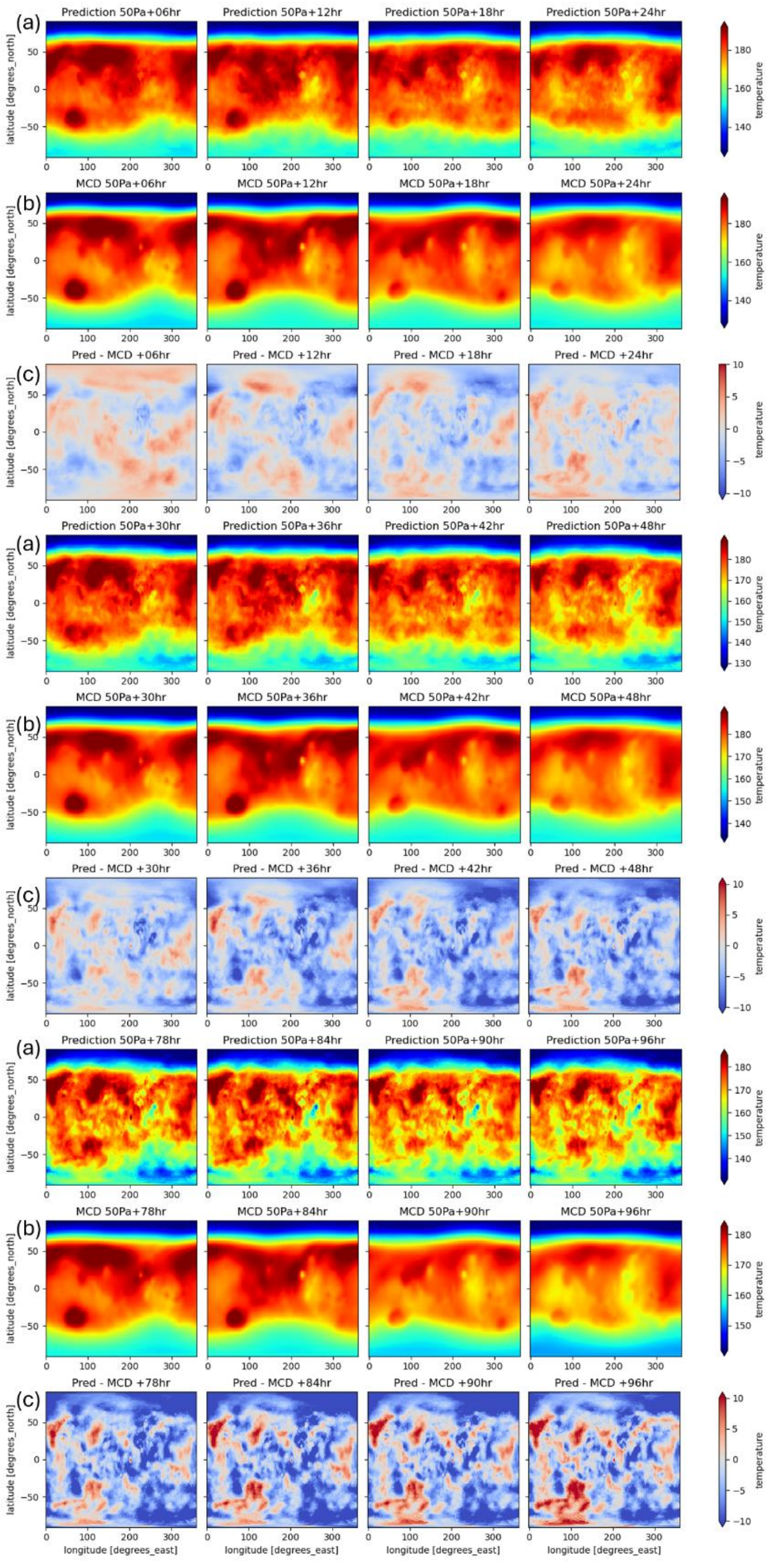

Figure S7: Temperature at 50Pa, initialized at Ls = 330, T00 and T06 for (a) MarsCast Prediction, b) MCD reference and (c) difference for +06 to +24 hours (top), +30 to +48 hours (middle), and +54 to 72 hours (bottom)

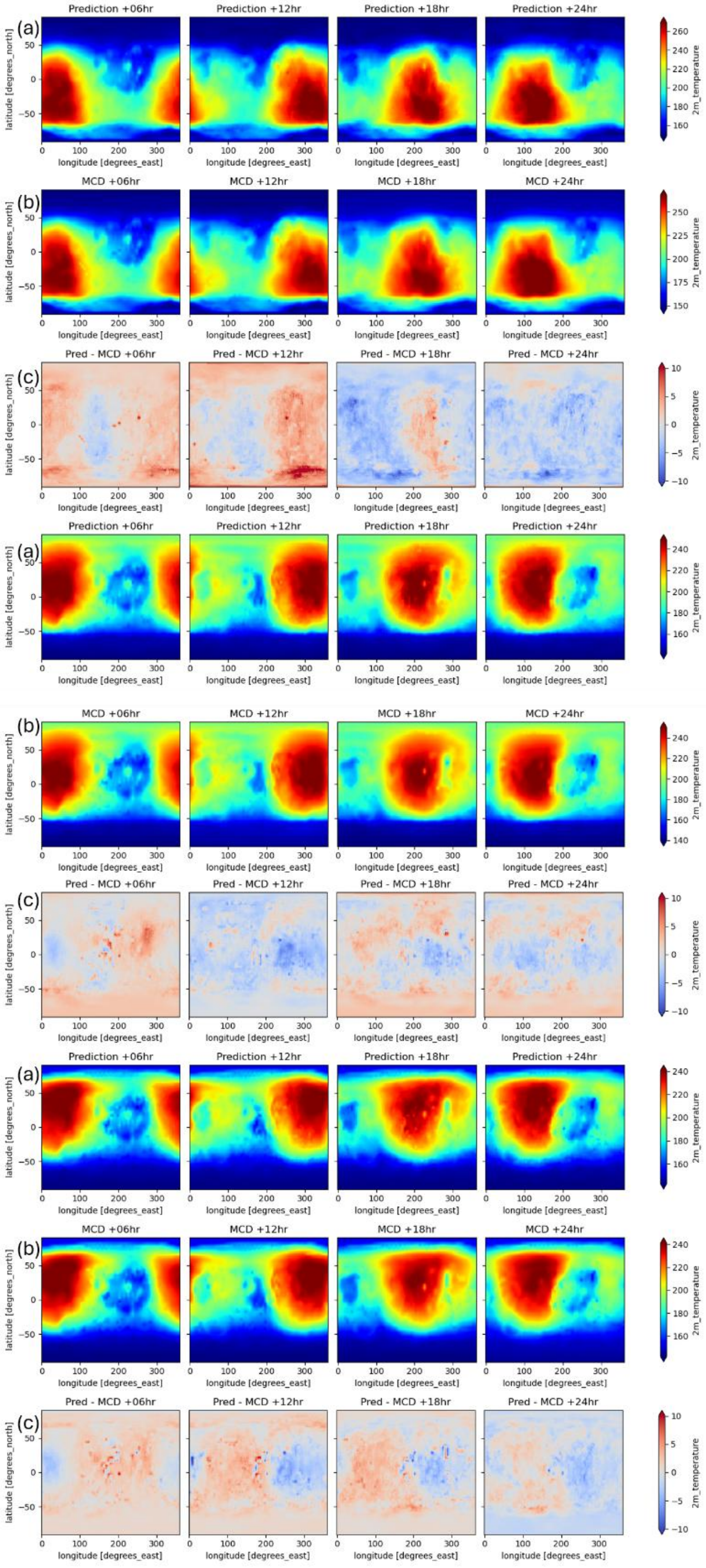

Figure S8: 2-m Surface Temperature (a) MarsCast Prediction, b) MCD reference and (c) difference for +06 to +24 hours, Ls 240/month 9/“Fall” (top), Ls 150/month 6/“Summer” (middle), and Ls 60/month 3/”Late Spring” (bottom).

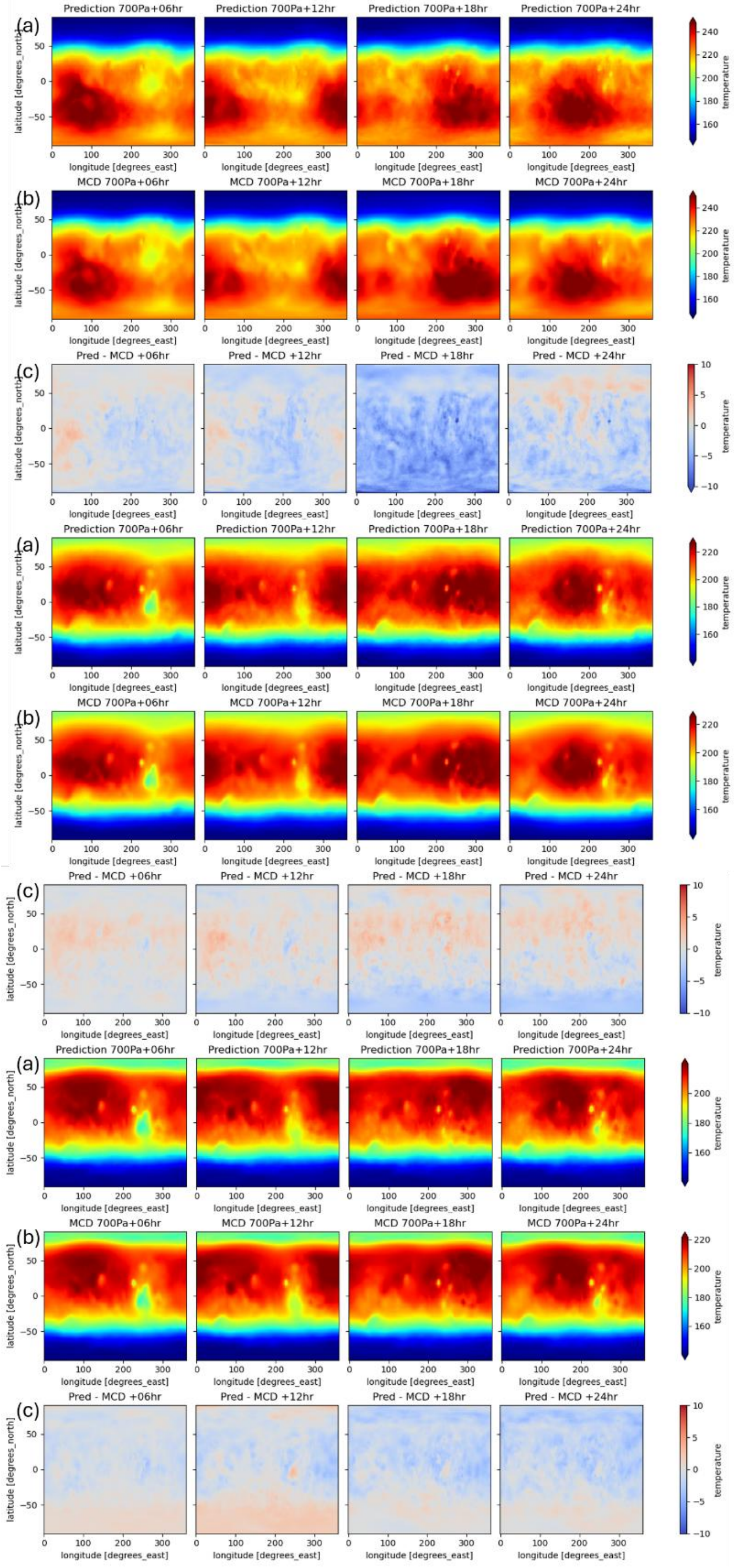

Figure S9: Temperature at 700Pa (a) MarsCast Prediction, b) MCD reference and (c) difference for +06 to +24 hours, Ls 240/month 9/"Fall" (top), Ls 150/month 6/"Summer" (middle), and Ls 60/month 3/"Late Spring" (bottom).

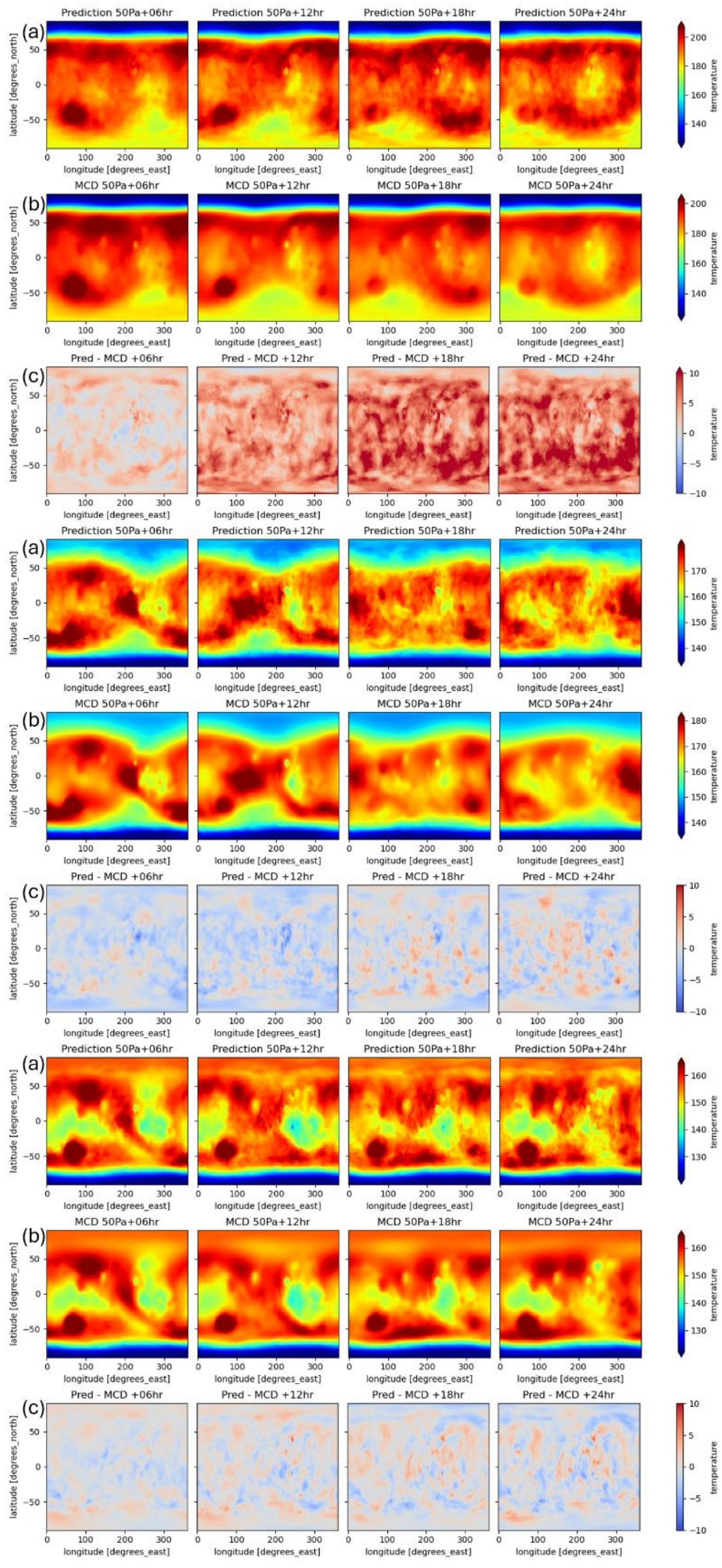

Figure S10: Temperature at 50Pa (a) MarsCast Prediction, b) MCD reference and (c) difference for +06 to +24 hours, Ls 240/month 9/“Fall” (top), Ls 150/month 6/“Summer” (middle), and Ls 60/month 3/”Late Spring” (bottom).

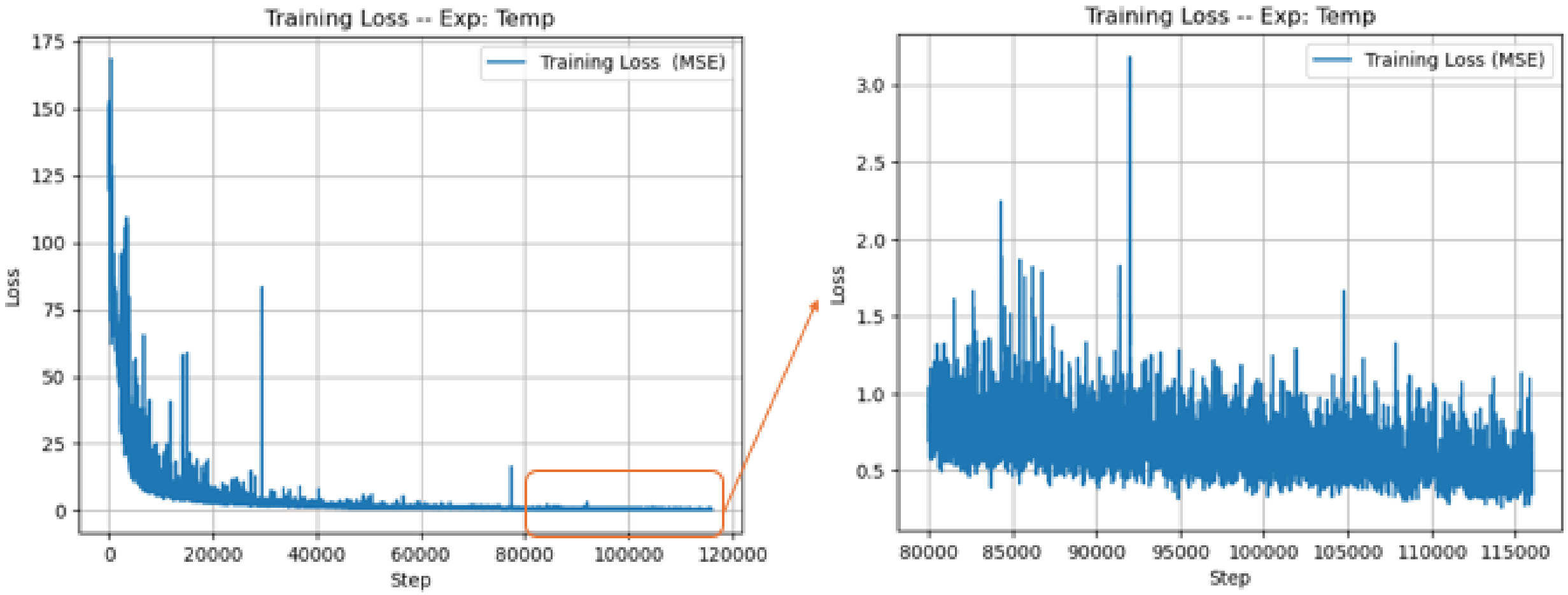

Figure S11: Training loss (MSE) values by steps (~120 steps per epoch). Training ran for 1,000 epochs and showed marked improvement for the first 300 epochs (~40k steps), and continued to show improvement even beyond 700 epochs (~80k steps).

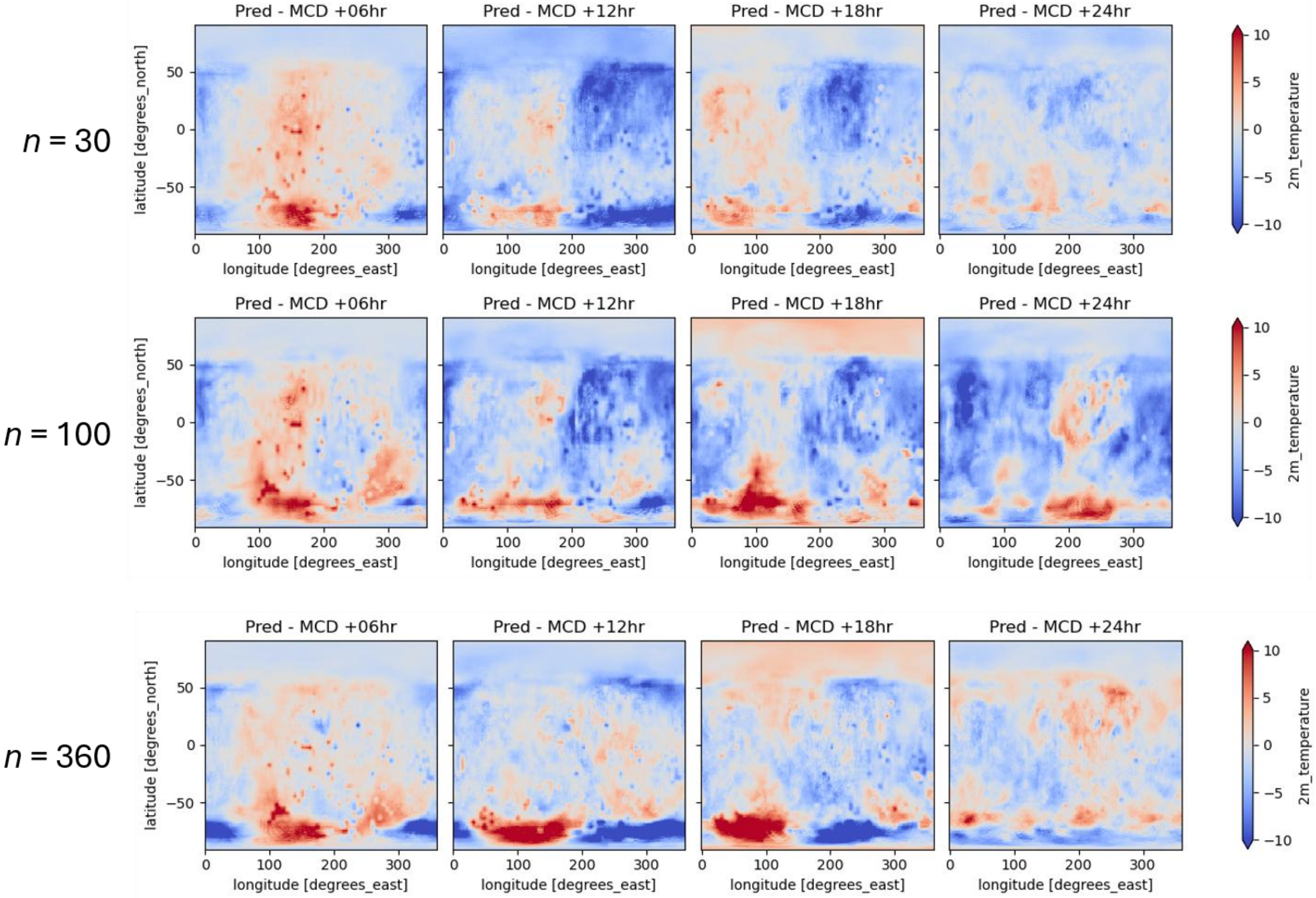

Figure S12. Difference between MarsCast and MCP (reference) values of the first 24-hour forecast (6 hour time steps), for the training sample sizes 30 (top), 100 (middle), and bottom (bottom). In each case fine-tuning was allowed to go to 100 epochs regardless of sample size. This likely did not allow enough time for the model to converge with greater amounts of data. However, the results with just 30 days of data and less fine-tuning were sufficient to demonstrate the capability of fine-tuning and Earth based AI weather model with Mars data.